\documentclass{goeproc}

\newcommand{\arcsec}{{\hbox{$^{\prime\prime}$}}}
\newcommand{\farcs}{\hbox{$.\!^{\prime\prime}$}}

\begin{document}

\title{On the inhomogeneities of the sunspot penumbra}

\author[1,*]{R.~Schlichenmaier}
\author[2]{D.A.N. M\"uller}
\author[1,3]{C. Beck}

\affil[1]{Kiepenheuer-Institut f\"ur Sonnenphysik, Freiburg, Germany}
\affil[2]{European Space Agency, c/o NASA Goddard Space Flight Center, Greenbelt, MD, USA}
\affil[3]{Instituto de Astrof\'{i}sica de Canarias, La Laguna, Tenerife, Spain}
\affil[*]{\textit{Email:} schliche@kis.uni-freiburg.de}

\runningtitle{On inhomogeneities of the sunspot penumbra}
\runningauthor{Schlichenmaier et al.}

\firstpage{1}

\maketitle

\begin{abstract}
The penumbra is ideally suited to challenge our understanding of magnetohydrodynamics. 
The energy transport takes place as magnetoconvection in inclined magnetic fields under 
the effect of strong radiative cooling at the surface. The relevant processes happen 
at small spatial scales. In this contribution we describe and elaborate on these 
small-scale inhomogeneities of a sunspot penumbra. We describe the penumbral 
properties inferred from imaging, spectroscopic and spectropolarimetric data, and 
discuss the question of how these observations can be understood in terms of 
proposed models and theoretical concepts.
\end{abstract}

\section{Introduction}

Almost hundred years ago, \citet{hale1908a, hale1908b} performed the first spectropolarimetric 
measurements on the Sun and discovered that sunspots are manifestations of magnetic fields 
with strengths of up to 3000 G in their centers. He proposed the tornado theory to 
explain both the darkness of a spot by dust that is whirled up into the solar atmosphere 
and the magnetic field by the circular current by electrons. Investigating Doppler shifts 
in sunspots, \citet{evershed1909} measured radial rather than circular movements in sunspot 
penumbrae and concluded that his findings were ''entirely out of harmony with the splendid 
discovery of the Zeeman effect in sun-spots, made by Prof. Hale."

Until today, we lack a complete understanding of sunspots and penumbrae, but it seems 
obvious that the Evershed flow plays an essential role in the latter. The presence of 
the Evershed flow is intimately linked to the mere existence of the penumbra as  
demonstrated by the observations of \citet{leka+skumanich1998}. The filamentary bright 
and dark structure should be related to the flow field, in a similar way as the granular 
pattern of the quiet Sun is related to the granular flow field of hot up- and cool downflows.

The magnetic field can -- in principle -- be inferred from the spectropolarimetric 
imprints of the Zeeman effect on photospheric absorption lines. However, this leads to 
unambiguous results only if the magnetic and velocity fields are homogeneous along the 
line-of-sight. Considerable complications in interpreting the line profiles arise if 
gradients or discontinuities are present in the volume of the solar atmosphere which is 
sampled by one resolution element. Such variations may be present laterally due to 
insufficient spatial resolution, but also along the line-of-sight from the depth layers 
that contribute to the observed line profile. In this contribution, we make the case that 
the latter must be assumed to reconstruct some of the observed profile asymmetries. 
We summarize our knowledge of the inhomogeneities in the magnetic and velocity field of 
a sunspot penumbra, aiming at an understanding of radiative magneto-convection in 
inclined magnetic fields.

\section{Penumbral properties}

The photospheric penumbral properties may be characterized by three different types of 
measurements: (1) imaging, (2) spectroscopy of lines that do not show the Zeeman effect, 
and (3) spectropolarimetric measurements of Zeeman sensitive lines.

\subsection{Imaging} 
At a spatial resolution of $1\arcsec$ or worse, the penumbra appears as a gray ring that 
surrounds the umbra. The radial bright and dark filaments only become apparent at a spatial 
resolution of about $0.5\arcsec$. At a spatial resolution of about $0\farcs2$ bright 
filaments show internal intensity variations: In the inner penumbra, predominantly on 
the center side, bright filaments show a dark elongated core \citep{scharmer+etal2002, 
suetterlin+bellot+schlichenmaier2004, langhans+etal2007}. Such dark-cored bright 
filaments have a width of some $0\farcs2$.

\subsection{Spectroscopy} 
As for the intensity, the length scale of variations in a Doppler map decreases with 
improving spatial resolution. At $1\arcsec$ no flow filaments are visible and the 
Doppler shifts only show a transition from red shifts on the limb side to blue shifts 
on the center side. To infer the flow vector at this spatial resolution, azimuthal cuts 
at constant distance from the spot center are constructed. The azimuthal mean reflects 
the vertical velocity component, while the amplitude of the variation measures the 
horizontal component which is known to be dominant. It is found that the flow field 
has an upward component in the inner and a downward component in the outer penumbra 
\citep{schlichenmaier+schmidt2000, schmidt+schlichenmaier2000}.

A filamentary structure becomes visible in velocity maps of the penumbra if the spatial 
resolution is of the order of $0\farcs5$ or better. However, flow filaments are not 
always co-spatial with intensity filaments \citep{tritschler+etal2004, 
schlichenmaier+bellot+tritschler2005}. Individual flow and intensity filaments typically 
have a joined starting point in the inner penumbra. Yet, further outwards the flow 
filament is not always co-spatial with a bright filament.

The Evershed flow in the penumbra is characterized not only by a line shift, but also 
by a line asymmetry. The line asymmetry is such that the wing is more strongly shifted 
than the core \citep[][]{bumba1960,  schroeter1965a, wiehr+etal1984, degenhardt1993, 
wiehr1995}. The asymmetry decreases with formation height of the spectral line 
\citep{degenhardt+wiehr1994, balthasar+schmidt+wiehr1997}. Studying the line asymmetry, 
i.e., the shape of the bisector, one can show that the measurements are consistent 
with the assumption that the flow is concentrated in the deepest photospheric layers 
\citep{maltby1964}. Line asymmetries could in principle be assigned to laterally 
separated unresolved components \citep{schroeter1965b}, but as they are still present 
in high spatial resolution observations down to $0\farcs3$ 
\citep[see][]{bellot+langhans+schlichenmaier2005}, the line asymmetries can only be 
explained by velocity gradients or discontinuities along the line-of-sight 
\citep{schlichenmaier+bellot+tritschler2004, bellot+schlichenmaier+tritschler2006}. 
Therefore, the measurements seem to require a flow field that is concentrated in the 
deepest layers of the photosphere beneath $\tau\approx 0.1$.

Spectroscopy at $0\farcs3$ spatial resolution resolves dark-cored bright filaments. 
\citet{bellot+langhans+schlichenmaier2005} found that the dark cores are associated with 
strong predominantly horizontal flows, while the lateral bright edges are less Doppler 
shifted. Yet the lateral bright edges are still more shifted than the dark surroundings 
in which the dark-cored penumbral filaments are often embedded in the inner penumbra.
The inner end of a dark-cored bright filament is typically bright and is associated 
with an upflow. \citet{rimmele+marino2006} find that these hot upflows have a diameter 
of about $0\farcs3$ and turn horizontal within $1\arcsec$. They continue in the dark core 
of a bright filament.

\subsection{Magneto-convetive modelling}
The first attempt to describe quantitatively (i) the dynamic evolution of penumbral 
filaments and (ii) the Evershed effect consisted of a 1D (thin) magnetic flux tube that 
evolved in a 2D time-independent background \citep{schlichenmaier+jahn+schmidt1998a, 
schlichenmaier+jahn+schmidt1998b}. Such tubes develop a flow that brings up hot plasma 
into the photosphere which radiatively cools as it flows outwards, yielding a flow 
topology that is consistent with the findings by \citet{rimmele+marino2006}. As long 
as the plasma is hot, the magnetic field strength is reduced and can be weaker by 
$10^3$ G as compared to the background. In this framework, magneto-convective instabilities 
cause solutions in the form of sea serpents \citep{schlichenmaier2002}, in which the 
hot upflow is followed by a cool downflow further outwards. Such a flow arch is 
conceptually very similar to the time-independent siphon flows 
\citep[e.g.,][]{degenhardt1991, thomas+montesinos1993}. In both cases a pressure gradient 
accelerates the flow. The pressure gradient in the moving tube model is sustained by 
radiative cooling between the inner hot footpoint (high gas pressure) and the cool 
flow channel further outwards (low gas pressure). This configuration is a natural 
consequence of magneto-convective instabilities and radiative cooling. 

In contrast, \citet{weiss+etal2004} have proposed that magnetic field lines associated 
with the downflows in the outer penumbra are pumped downwards by small-scale granular 
convection outside the sunspot. They demonstrate that turbulent transport of mean 
magnetic fields by convective motions, i.e., magnetic pumping, is a robust phenomenon. 
However, in this scenario it remains obscure how the magnetized penumbra with supposedly 
coherent flows and field lines is connected to the turbulent convection that is needed 
to produce topological pumping effects. It seems more promising to perform 3D simulations 
of inclined magnetoconvection including the effects of radiation. First results have 
been presented by \citet{heinemann+etal2006} which confirm the flow topology that we 
describe above of a hot upflow followed by a cooler downflow (cf. their Fig. 6). Yet, in 
these simulations the flow is not strictly radial, but has a small azimuthal component 
towards the lateral edges of the filament. Eventually such simulations will clarify whether 
or not the energy transport takes place due to convection in field free gaps as it was 
suggested recently \citep{spruit+scharmer2006, scharmer+spruit2006} or by flows along 
magnetic tubes as suggested by the moving tube model. In any model, the magnetic field 
strength in bright penumbral features is certainly less than 1000 G, i.e., much smaller 
than the average surrounding magnetic field strength. In accordance with the latter 
expectation, \citet{bellot+langhans+schlichenmaier2005} measure that dark-cored 
filaments have smaller magnetic field strengths than the surroundings of the filaments. 
They use the Zeeman sensitive line Fe\,{\sc ii}\,6149\,nm and even find a tendency that the 
strength is slightly larger in dark cores than in the bright lateral edges, as one 
expects due to contraction of the plasma during the cooling phase.

\begin{figure}[t]
\parbox[b]{9.2cm}{\includegraphics[width=9cm]{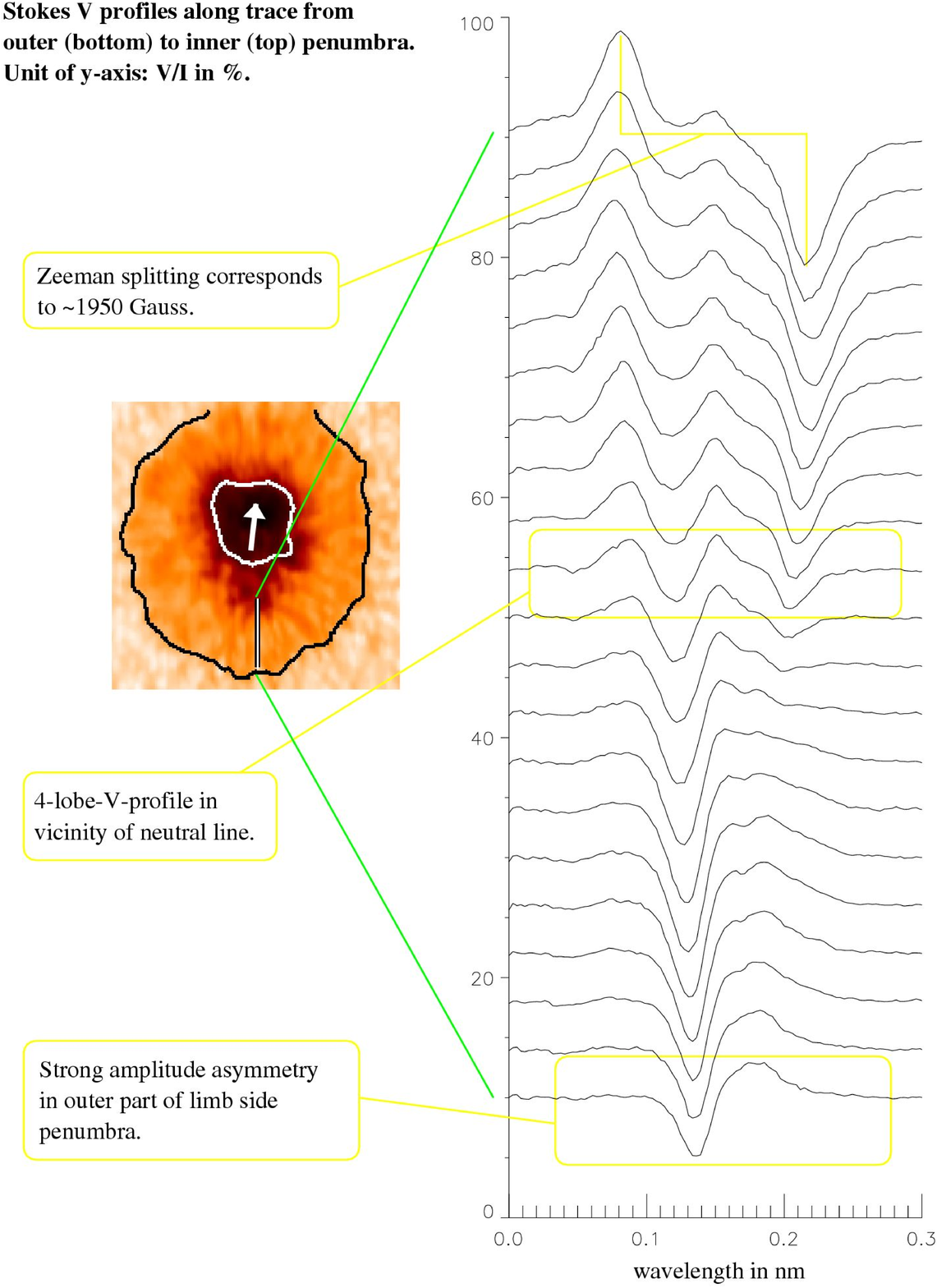}}
\hfill
\parbox[b]{3.5cm}{
\caption[]{\label{fig:4-lobe-profile}\sloppy
Stokes-V profiles of Fe\,{\sc i}\, 1564.8\,nm along a radial cut for the limb-side 
penumbra of a spot at 30$^\circ$ heliocentric angle. From the inner to the outer 
penumbra, the V-profiles undergo a transition from one polarity to the other. 
In the central penumbra, along the so-called magnetic neutral line, both 
polarities seem to be present. This is clear observational evidence for an 
inhomogeneous magnetic field. Such a 4-lobe-profile needs at least 
two different orientations of the magnetic field within one resolution 
element.\vspace{2.0cm}}}
\end{figure}

\begin{figure}[t]
%\parbox[b]{10cm}{
\center
\includegraphics[width=11cm]{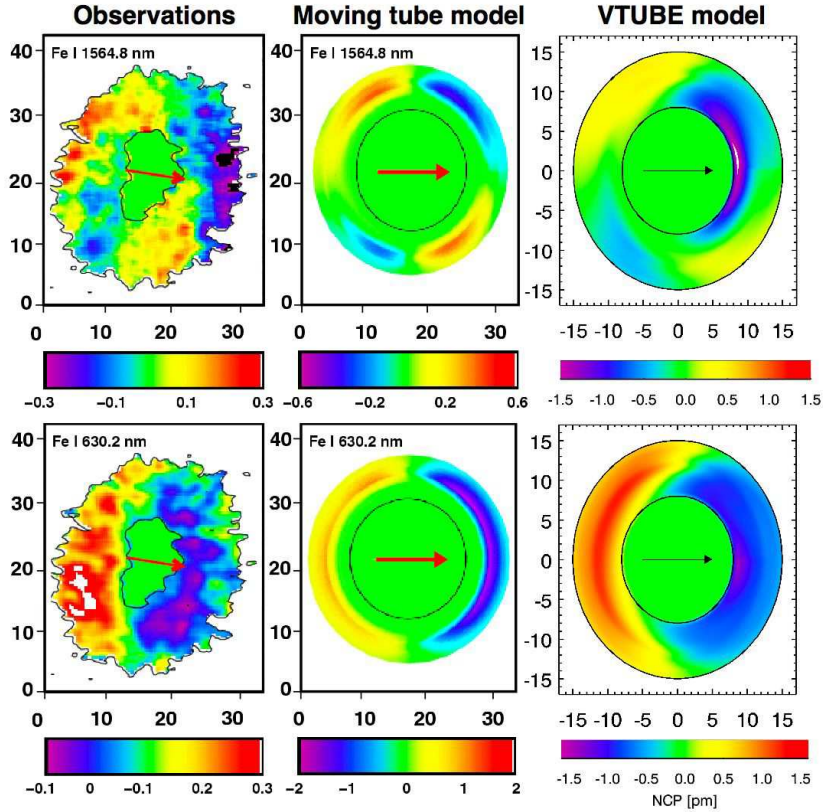}
%}
%\hfill
%\parbox[b]{3cm}{
\caption[]{\label{fig:ncp}\sloppy Observed and synthesized sunspot maps of net 
circular polarization for Fe\,{\sc i}\,15648\,nm (top row) and Fe\,{\sc i}\,630.2\,nm 
(bottom row), respectively. The synthesized maps in the middle column are based on 
a snapshot from the moving tube model \citep{mueller+etal2002}, and the maps in the 
right colum are based on the VTUBE tool (see text), where inversion results are 
used. The spot is at a heliocentric angle of 30$^\circ$ and the arrows point 
towards disk center.
}
%}
\end{figure}

\begin{figure}[t]
\parbox{6.7cm}{\includegraphics[width=6.5cm]{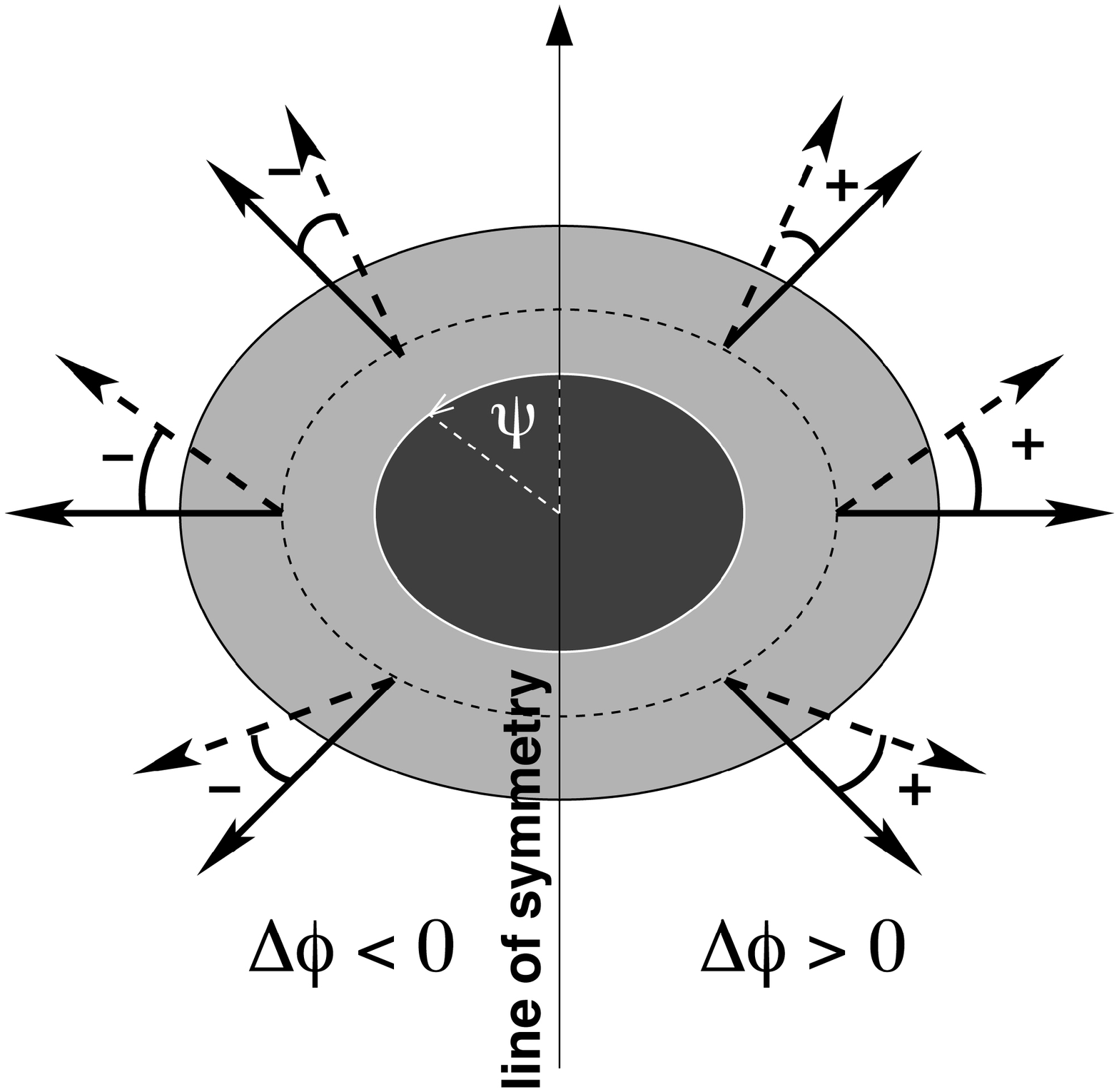}}
\hfill
\parbox{6cm}{
\caption[]{\label{fig:angles}\sloppy Difference in the azimuth, $\triangle\phi$, 
of horizontal and inclined magnetic field when viewed off disk center. The sign 
of $\triangle\phi$ changes from one side of the penumbra to the other. The dashed 
vectors represent the less inclined background magnetic field being at rest, and 
the solid vectors represent the horizontal magnetic components that is coaligned 
with the flow.
}}
\end{figure}

\subsection{Spectropolarimetry}

Spectroscopic measurements establish the inhomogeneous flow field within the sunspot 
penumbra. Direct observational evidence for an inhomogeneous magnetic field consists of 
4-lobe V-profiles as presented in Fig.~\ref{fig:4-lobe-profile}: At the location within 
the penumbra where the viewing angle to the average magnetic field lines is perpendicular, 
i.e., in the vicinity of the magnetic neutral line, one expects vanishing V-signal if 
the field were homogeneous. V-profiles with up to four lobes, as observed, imply that 
at least two directions (components) of the magnetic field must be present in the 
resolution element. Another strong indication of non-trivial topologies of the velocity 
and magnetic field are given by the observations of amplitude and area asymmetries of 
Stokes-V \citep[e.g.,][]{sanchez+lites1992, schlichenmaier+collados2002}. These findings 
motivate that reconstructions of the (magnetic and velocity) field topology should 
consider at least two components or strong gradients or discontinuities along the 
line-of-sight. Such attempts have been made using inversion techniques 
\citep[e.g.,][]{bellot+balthasar+collados2004, borrero+etal2006}. In such inversions, 
synthetic Stokes parameters resulting from polarized radiative transfer based on model 
atmospheres are compared with observed profiles of the four Stokes parameters.

The assumption of two interlaced components with constant velocity and magnetic fields 
give reasonable fits to the observed profiles. Yet, the pure existence of net circular 
polarization, ${\cal N} = \int V(\lambda) \,{\rm d} \lambda$, proves that velocity gradients 
along the line-of-sight are present \citep{sanchez+lites1992, landolfi+landi1996}.
Sunspot maps of the net circular polarization are displayed in Fig.~\ref{fig:ncp}. 
The first row shows observed and synthesized maps of Fe\,{\sc i}\,15648\,nm, and the second 
row shows the same for Fe\,{\sc i}\,630.2\,nm. The observed maps of ${\cal N}$ (left column) 
are simultaneous measurements of the same sunspot at 30$^\circ$ heliocentric angle 
\citep{mueller+etal2006}. The symmetry properties of the two observed maps are different.
The middle column shows synthetic maps that are based on a snapshot of the moving 
tube model \citep{mueller+etal2002}. The synthetic maps reproduce the observations quite 
convincingly. The main ingredient for reproducing these symmetry properties are two 
components along the line-of-sight with the following properties (1) They have two 
different inclinations of the magnetic field and (2) they are Doppler shifted relative 
to each other \citep{schlichenmaier+etal2002}. Figure \ref{fig:angles} sketches how 
the two different inclinations of the magnetic field produce a sign change in the 
difference of the magnetic field azimuth, which is crucial to reconstruct the observed 
maps. It is important to realize that these maps cannot be reproduced with a mixture of 
two laterally separated (unresolved) components with each having constant magnetic and 
velocity fields along the line-of-sight! 

\section{A diagnostic tool: VTUBE}

In an attempt to bring together snapshots from simulations and results from inversions, 
a tool to diagnose spectropolarimetric measurements was developed \citep{mueller+etal2006}. 
We constructed a 3D geometric model (VTUBE) of a magnetic flux tube embedded in a 
background atmosphere that can be used to gain a better understanding of the different 
factors that determine $\cal N$ and its spatial variation within the penumbra. This model 
serves as the frontend for a radiative transfer code \citep[DIAMAG,][]{grossmann1994}. 
Combining the two, we can generate synthetic Stokes spectra for any spectral line and 
construct maps of suited diagnostic quantities, like $\cal N$-maps, for any desired 
axisymmetric magnetic field configuration and arbitrary properties of flux tubes being 
embedded in an arbitrary atmosphere. The model has been built to offer a high degree of 
versatility, e.g.\ the option to calculate several parallel rays along the line-of-sight 
that intersect the tube at different locations with arbitrary viewing angles. One can 
then average over these rays to model observations of flux tubes at different spatial 
resolutions and for different magnetic filling factors of the atmosphere. Furthermore, 
one can also take into account radial variations of the physical properties of the flux 
tube. Doing so, one can e.g.\ model the interface between the flux tube and its surroundings.

To generate a realistic $\cal N$-map with the VTUBE model, we extracted the radial 
dependence of azimuthally averaged quantities from results of inversions based on two 
components as mentioned above, thereby assuming that the penumbral fine structure is 
axially symmetric. From the two components, one is assigned to be the background component, 
and the other serves as the flow component. This is in accordance with inversion 
results with one component being essentially at rest and the other the other carrying 
the flow. While the magnetic field strength decreases with penumbral radius, the 
difference between the weaker flow component and the stronger one decreases outwards 
and vanishes at the outer penumbral edge. The flow velocity along the channel 
(flow component) increases from some 6 km/s at the inner to 8 km/s in the outer penumbra. 
The inclination of the flow changes gradually from an upflow with $20^\circ$ in the 
inner to a downflow with $-10^\circ$ in the outer penumbra. The resulting maps 
(right column in Fig.~\ref{fig:ncp}) compare well with the observed ones, but, especially 
in the radial dependence, differences are seen. Possibly this difference may be explained 
by recent results \citep{bellot+etal2003a, beck2006} that suggest that the background 
field is not at rest in the outermost penumbra, but shows flow velocities of up to 
1 km/s. This may modify the $\cal N$-map and may reconcile the model with the observations.

\section{Conclusions}

We have described the properties of a sunspot penumbra, including the topology of the 
magnetic and velocity field. The main conclusion is that the penumbra is inhomogeneous. 
In order to explain the mentioned spectroscopic and spectropolarimetric measurements 
at least strong gradients if not discontinuities must be present along the line-of-sight. 
We believe that low lying channels ($\tau > 0.1$) that carry a magnetized flow and are 
embedded in a magnetized background at rest must be assumed in order to understand the 
measurements, like the observed bisectors of unmagnetic lines, and the symmetry 
properties of net-circular-polarization maps. Yet, we cannot offer a complete picture as 
the spatial resolution is not yet good enough to present unambiguous model fits to the 
observed spectropolarimetric measurements. A great step forward may be accomplished by 
the recently launched satellite Hinode, as it will allow for spectropolarimetric 
observations at 0\farcs25. The best ground based observations today are not much better 
than 1\arcsec. At the same time, simulations of magnetoconvection in inclined magnetic 
field like the ones presented recently by \citet{heinemann+etal2006} will significantly 
enhance our understanding of the penumbral fine structure.

\end{document}